\documentclass{PoS}
\usepackage{epsfig}
\usepackage{wrapfig}

\title{Results from the MINOS Experiment and New MINOS+ Data}

\ShortTitle{MINOS and MINOS+ Results}

\author{\speaker{Anna Holin}%
        \thanks{On behalf of the MINOS and MINOS+ Collaborations}\\
       University College London\\
       E-mail: \email{anna.holin@ucl.ac.uk}}

\abstract{The MINOS experiment took data for seven years between May 2005 and April 2012. Since then it has been reborn as the new MINOS+ experiment in the upgraded medium energy NuMI beam and started taking data in September 2013. An update to the MINOS standard oscillations three-flavour disappearance analysis is presented which includes 28\% more atmospheric neutrino data. This combined three-flavour analysis calculates an atmospheric parameter best-fit point of $\Delta m_{32}^{2}=2.37^{+0.11}_{-0.07} \times 10^{-3}$~eV$^{2}$ and $\sin^{2}\theta_{23}=0.43^{+0.19}_{-0.05}$ for the inverted hierarchy, for which the MINOS fit shows a slight preference. A first look at the new MINOS+ beam data is presented. The new data is consistent with the combined three-flavour analysis. Finally, new MINOS results for the search for sterile neutrinos using neutrino disappearance are shown which cut out a significant amount of the allowed phase space for a sterile neutrino to exist.}

\FullConference{16th International Workshop on Neutrino Factories and Future Neutrino Beam Facilities - NUFACT2014,\\
  25 -30 August, 2014\\
  University of Glasgow, United Kingdom }

\begin{document}

\section{Introduction}
Over the past decade, MINOS has been successfully measuring fundamental neutrino oscillations parameters. Using the NuMI beam \cite{NuMI} and its two-detector design, a Near Detector (ND) at the beam source at Fermilab and a Far Detector (FD) 735~km away in the Soudan Mine in Minnesota \cite{detectorNIM}, MINOS was able to cancel out many systematic uncertainties, for example associated with Neutrino Flux \cite{PRD77}, and push the boundaries of the measurements that could be made \cite{Meas1}-\cite{Meas3}. 

MINOS took data between May 2005 and April 2012, mostly in the so-called Low Energy (LE) beam configuration, with the neutrino spectrum peaked at approximately 3~GeV, close to the standard 3-flavour oscillations maximum at about 1.6~GeV in the MINOS Far detector. This allowed MINOS to carry out precise measurements of the muon neutrino disappearance oscillations parameters. MINOS was then reborn as the ongoing MINOS+ experiment \cite{minosplusref}, and is now taking data in the NO$\nu$A-era NuMI beam, which started producing neutrinos in September 2013 and is running at a higher energy, optimal for the NO$\nu$A experiment \cite{NOvA}. This so-called Medium Energy (ME) beam is peaked at approximately 7~GeV, relatively far from the standard 3-flavour oscillations maximum, however, on the other hand, MINOS+ is currently the only neutrino experiment in the world seeing a wide-band on-axis beam, with a wider neutrino energy range than other accelerator neutrino experiments such as NO$\nu$A and T2K, and with significantly more events per unit of exposure due to being on-axis. MINOS+ has therefore a unique potential to look for new physics and exotic phenomena such as sterile neutrinos, Non-Standard Interactions (NSI) \cite{NSI-C}, and large extra dimensions. This proceedings paper summarises some MINOS and MINOS+ results as of October 2014. 

\subsection{The MINOS(+) Experiment}
The MINOS(+) experiment consists of two detectors, a Near and a Far Detector. The detectors are in the path of the NuMI neutrino beam. The ND is located at Fermilab, 1.04~km from the NuMI beam source, and its purpose is to measure the neutrino beam before oscillations have occured. The FD is located $\sim$700~m underground in the Soudan Mine, and measures the neutrino beam after the neutrinos have oscillated. 

In order to generate neutrinos, the NuMI beam sends bunches of 120~GeV protons onto a 1~m long graphite target. This generates hadrons which are focused by two magnetic horns and decay into secondaries and neutrinos in a 675~m long decay pipe. The secondaries are then removed from the beam by a hadron absorber at the end of the decay pipe and a ``muon shield'' consisting of 240~m of dolomite rock on the way to the MINOS ND. This leaves a neutrino beam that consists mostly of muon neutrinos (91.7\%), with an admixture of muon antineutrinos (7\%), and a small electron neutrino contamination (1.3\%). The NuMI beam can be modified to produce an antineutrino enhanced beam (rather than the standard neutrino-dominated beam) by reversing the horn current and thus focusing negative hadrons. 

\begin{figure}
\begin{centering}
    \includegraphics[width=0.47\textwidth]{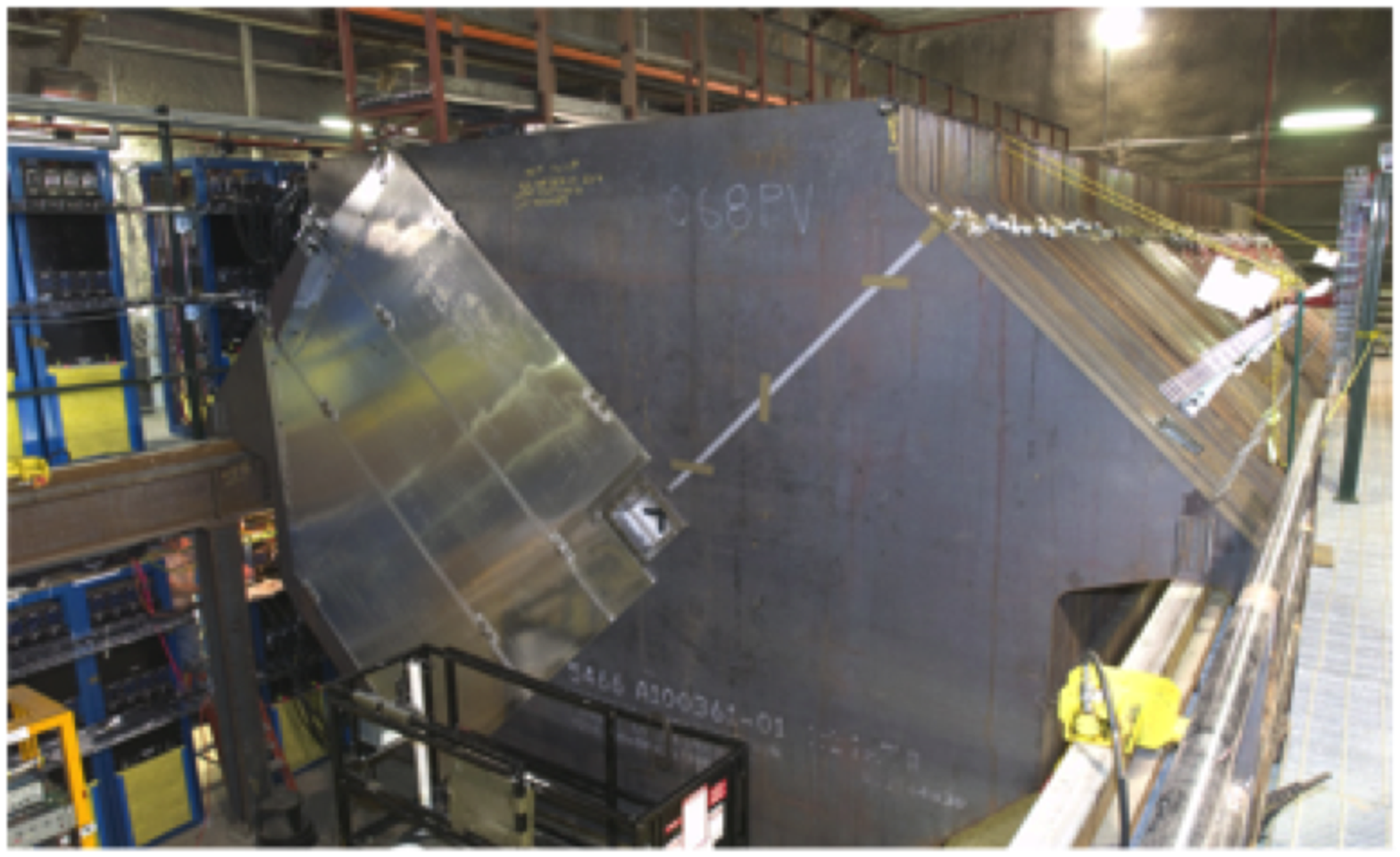}
    \includegraphics[width=0.47\textwidth]{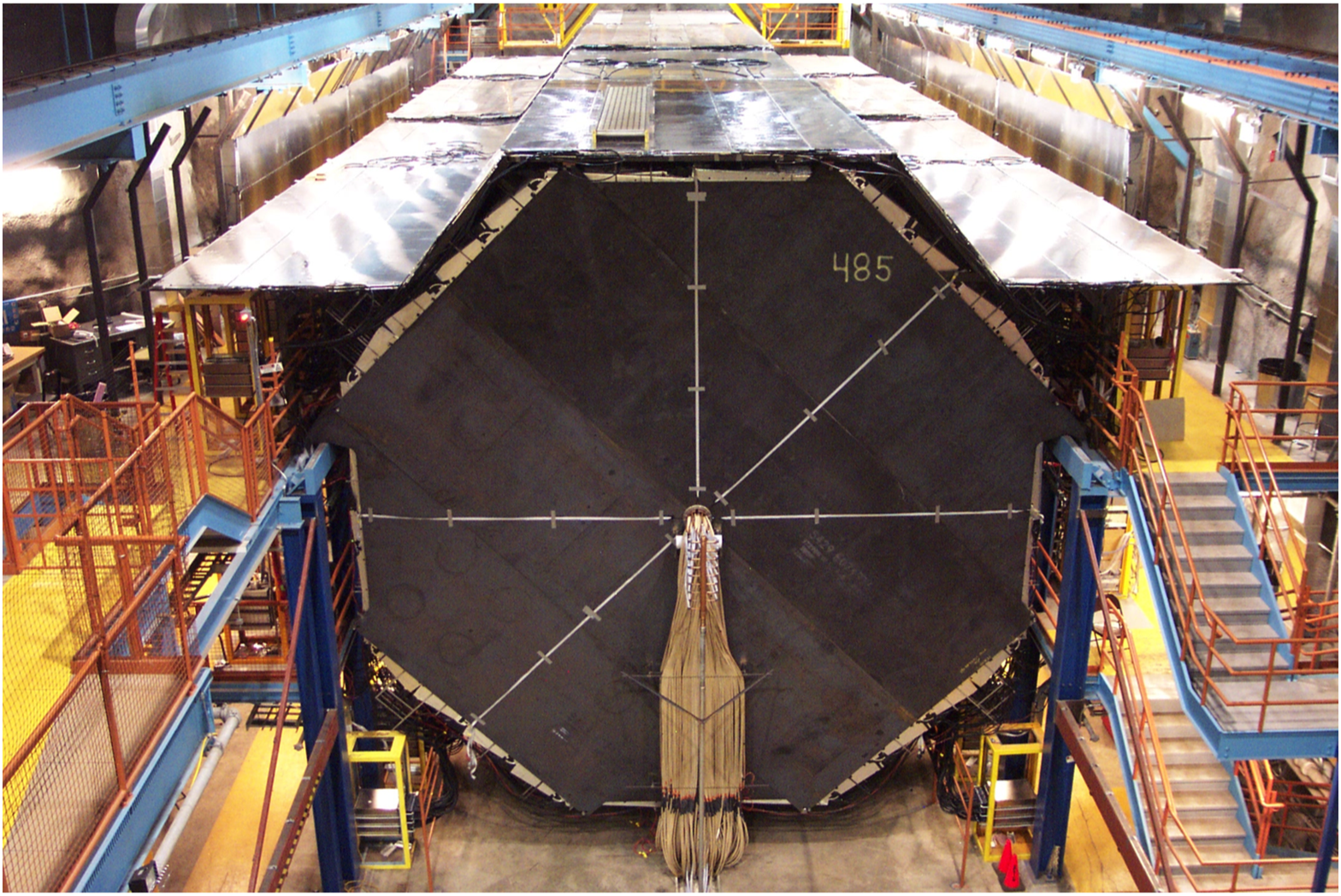}
    \caption{Pictures of the two MINOS+ Detectors. The Near Detector is smaller, at about 1~kT mass and 15~m length. The Far Detector is very large at 5.4~kTon mass and 30~m length, and has a veto shield above it to identify cosmic ray muons. }
\label{fig:detpics}
\end{centering}
\end{figure}

Both MINOS(+) detectors are large magnetized iron scintillator tracking calorimeters optimised to measure muon tracks from muon neutrino charged current (CC) interactions and to separate muon neutrinos from antineutrinos based on track curvature. They are built out of alternate planes made of steel and scintillator. The scintillator planes are attached to the steel planes and consist of long scintillator strips with embeded wavelength shifting fibers that are read out by photomultiplier tubes. Subsequent scintillator planes are positioned orthogonally to each other for 3-D reconstruction of events. The steel planes serve as a large detector mass to increase the likelihood of neutrinos interacting in the detector. Figure \ref{fig:detpics} shows pictures of the MINOS Near and Far detectors. The ND is 15~m long, 3.8~m tall and 4.8~m wide. It weighs approximately 1~kT and has a ``calorimeter'' region where each steel plane is instrumented with scintillator, and a ``spectrometer'' region where only every fifth plane is instrumented; there are 282 steel planes, whether instrumented or not, in total. The FD is 30~m long, 8~m wide and 8~m tall. It weighs 5.4~kT and consists of 486 fully instrumented steel/scintillator planes. It also has a veto shield to identify cosmic ray muons.  

\subsection{Neutrino Oscillations}
There are three known active neutrino flavours, the electron, muon, and tau neutrino ($\nu_{e}$,$\nu_{\mu}$,$\nu_{\tau}$). Each neutrino also has a corresponding antineutrino.  Neutrinos interact only via the weak force (and gravity), and therefore very rarely, and have been observed to oscillate between the flavours as they travel. Neutrino oscillations can be described by Equation \ref{equ:simplenuosc}:
\begin{equation}
\left(\begin{array}{c}\nu_{e}\\\nu_{\mu}\\\nu_{\tau}\end{array}\right)=U \left(\begin{array}{c}\nu_{1}\\\nu_{2}\\\nu_{3}\end{array}\right)
\label{equ:simplenuosc}
\end{equation}

where U is the Pontecorvo-Maki-Nakagava-Sakata (PMNS) matrix \cite{PMNSmatrix1}-\cite{PMNSmatrix3} which has a parametrisation of the form:
\scriptsize
\begin{equation}
U=\left(\begin{array}{ccc}c_{12}c_{13}&s_{12}c_{13}&s_{13}e^{-i\delta}\\-s_{12}c_{23}-c_{12}s_{23}s_{13}e^{i\delta}&c_{12}c_{23}-s_{12}s_{23}s_{13}e^{i\delta}&s_{23}c_{13}\\s_{12}s_{23}-c_{12}c_{23}s_{13}e^{i\delta}&-c_{12}s_{23}-s_{12}c_{23}s_{13}e^{i\delta}&c_{23}c_{13}\end{array}\right)
\label{equ:pmnsmatrix}
\end{equation}
\normalsize

In the above, $\nu_{e,\mu,\tau}$ are the neutrino flavour eigenstates, $\nu_{1,2,3}$ are the neutrino mass eigenstates, $c_{ij}=\cos\theta_{ij}$, $s_{ij}=\sin\theta_{ij}$, with $\theta_{ij}$ being the mixing angle, and $\delta$ is the charge-parity (CP) violating phase. 

For a two-neutrino approximation where the difference in mass between two of the mass states is very small, the probability that a muon neutrino will have oscillated to a tau neutrino as it propagates is to first order given by:
\begin{equation}
P_{\nu_{\mu} \rightarrow \nu_{\tau}} = \sin^{2}2\theta \sin^{2}\biggl(\frac{1.27\Delta m^{2} L}{E}\biggr)
\label{equ:twoflavour2}
\end{equation}

where  $L$ is the baseline in km, $E$ is the neutrino energy in GeV, and $\Delta m^{2}$ is the mass squared difference between the two effective flavour states in $eV^{2}/c^{4}$.  Three-flavour neutrino oscillations are actually parametrised using two mass squared differences - $\Delta m_{21}^{2}$ (solar sector) and $\Delta m_{32}^{2}$ (atmospheric sector), three mixing angles - $\theta_{12}$, $\theta_{13}$, and $\theta_{23}$, and a Charge Parity violating phase - $\delta$. As the $\theta_{13}$ angle has now been measured by the Daya Bay \cite{dayabay}, Double Chooz \cite{doublechooz} and RENO \cite{renoexp} experiments to be $\approx9^{\circ}$, it can no longer be treated as negligible, and the full three-flavour framework has been adopted by MINOS for oscillation measurements. Currently the outstanding unknowns in the standard neutrino oscillations framework are the neutrino mass hierarchy (whether the mass state $m_{3}$ is heavier or lighter than the $m_{1}$-$m_{2}$ mass state ``doublet''), the CP violating phase $\delta$, and the precise value of the mixing angle $\theta_{23}$, including the octant of the $\sin^{2}\theta_{23}$ term.

\subsection{Sterile Neutrinos}
Though LEP has measured the number of light neutrinos to be 2.984$\pm$0.008 \cite{lep} \footnote{This number has since been revised slightly downward to 2.92$\pm$0.05 \cite{pdg}.} there is still the possibility that there are sterile neutrinos that do not interact via the weak force. There could be a fourth (or more) neutrino mass state that oscillates together with the standard active neutrinos but cannot be detected via normal charged current (CC) and neutral current (NC) interactions. For example, assuming a simple 3+1 model, this could lead to the observation of additional neutrino disappearance depending on the value of the third mass squared difference term $\Delta m_{43}^{2}$.  There would also be additional mixing angles $\theta_{14}$, $\theta_{24}$ and $\theta_{34}$.

Sterile neutrinos could explain some of the anomalies seen in the neutrino sector such as the LSND \cite{LSND} and MiniBooNE \cite{MiniB} results, but so far the evidence has been inconclusive with tension between appearance and disappearance results.

\section{Results}

\subsection{Standard Oscillations Three-Flavour Results}
New results are presented for the combined three-flavour analysis using $\nu_{\mu}$ disappearance data, $\nu_{e}$ appearance data, and an increased atmospheric event sample of 48.7~kT-years. As in the previous results from this analysis \cite{Meas3}, the full MINOS beam data set is used with a total exposure of 10.71$\times10^{20}$~protons-on-target (POT) of neutrino-dominated mode beam data, and 3.36$\times10^{20}$~POT of antineutrino-enhanced beam mode data. The larger atmospheric event sample used here corresponds to a 28\% increase on the previous analysis, which used 37.88~kT-years of data.

\begin{figure}
\begin{centering}
    \includegraphics[width=0.8\textwidth]{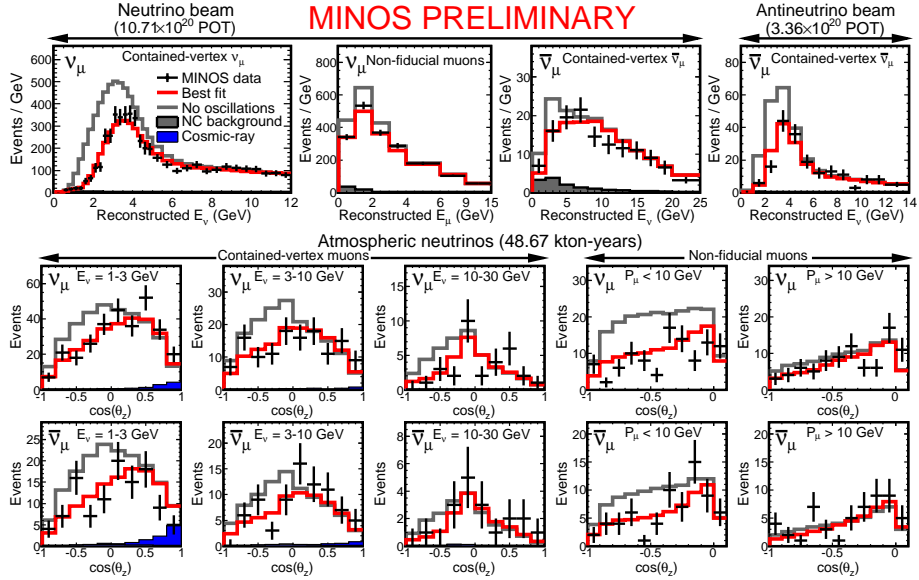}
    \caption{Main Samples used in Combined Analysis. MINOS beam data are shown in the top row, and the combined MINOS and MINOS+ atmospheric data for contained-vertex and anti-fiducial samples of neutrinos and antineutrinos are shown in the middle and bottom rows. The distributions of atmospheric neutrinos are plotted as a function of zenith angle and separated into bins of reconstructed neutrino energy. The observed data (black points) are compared with the prediction for no oscillations in gray and the best fit three-flavour oscillations in red. The cosmic ray background is shown in blue while the NC background is shown in gray.   }
\label{fig:spectra}
\end{centering}
\end{figure}

For the $\nu_{\mu}$ beam data, CC interactions are selected by requiring that there be a muon track and applying a k-Nearest-Neighbor algorithm \cite{kNN} using various muon track variables related to energy deposition and topology of the track. Both a contained-vertex (fiducial volume) sample is selected for the beam muon neutrino events and a non-fiducial muon sample, with vertices outside the detector. A special selection is also applied to select a sample of muon antineutrino events in the neutrino beam, and to select muon antineutrinos in the antineutrino-enhanced beam.  For the $\nu_{e}$ beam data, CC events are selected using a library-event-matching (LEM) algorithm, which compares shower events to large samples of tens of millions of NC background and CC-$\nu_{e}$ signal events in order to determine whether a given event is CC-$\nu_{e}$-like \cite{Meas2}. To include atmospheric neutrinos and distinguish them from cosmic ray background events in the combined analysis, several different categories of events are analysed (see \cite{atm} for detail). The two main sample categories are contained-vertex events and non-fiducial (upward going or horizontal muon track) events. The FD veto-shield helps in reducing the background cosmic ray muon events by checking for coincidences between hits in the shield and the selected tracks. The contained-vertex and non-fiducial atmospheric neutrino samples are further split into muon neutrino and antineutrino samples using track curvature, and into different energy slices. Finally, an atmospheric shower selection is also included in the combined analysis. A summary of the muon neutrino spectra used in this analysis is shown in Fig. \ref{fig:spectra}. 

In the final fit, the $\Delta m_{32}^{2}$, $\theta_{23}$, and $\delta_{CP}$ parameters are left unconstrained. $\theta_{13}$ is fit as a nuisance parameter and constrained by the reactor results ($\sin^{2}\theta_{13}=0.0242\pm0.0025$). The solar mixing parameters are fixed to $\Delta m_{21}^{2}$=7.54$\times10^{-5}$~eV$^{2}$ and $\sin^{2}\theta_{12}$=0.307. Systematic uncertainties are taken care of by the inclusion of 32 different nuisance parameters in the fit. Due to the separating of the neutrino and antineutrino samples, the combined analysis is sensitive to the neutrino mass hierarchy, the value of $\delta_{CP}$, the octant of $\theta_{23}$, and the value of $\theta_{13}$. The calculations of oscillation probabilities in this analysis use the full three-flavour PMNS matrix calculations, with the inclusion of the MSW \cite{MSW}-\cite{MSW2} effect. For the beam events, the ND is used to predict the FD event spectra.

\begin{figure}
\begin{centering}
    \includegraphics[width=0.7\textwidth]{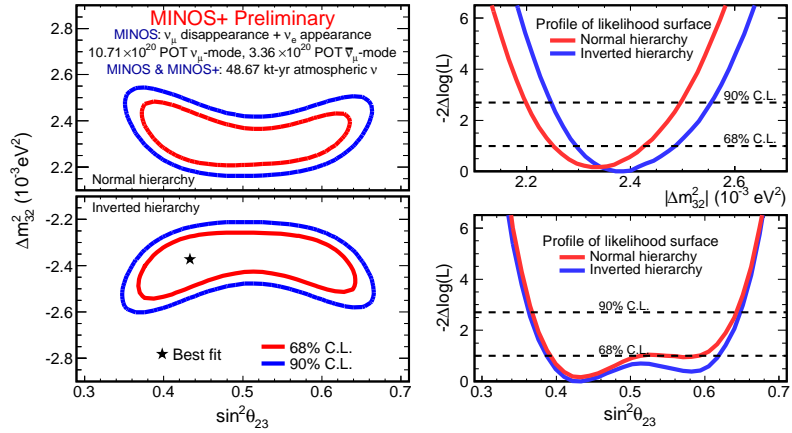}
    \caption{Contours and Profiles from Fit to 48.67~kT-years of atmospheric data combined with disappearance and appearance data from the MINOS beam. Left panels shows 68\% and 90\% confidence limits in ($\Delta m^{2}_{32}$, $\sin^{2}\theta_{23}$) calculated for normal hierarchy (top) and inverted hierarchy (bottom). Right panels show log-likelihood profiles for each hierarchy plotted for $\Delta{m}^{2}$ (top right) and $\sin^{2}\theta_{23}$ (bottom right). The best fit is indicated by the star and occurs in the inverted hierarchy at $\Delta m^{2}_{32} = -2.37 \times 10^{-3}$\ eV$^{2}$ and $\sin^{2}\theta_{23} = 0.43$.   }
\label{fig:contours}
\end{centering}
\end{figure}

\begin{figure}
\begin{centering}
    \includegraphics[width=0.47\textwidth]{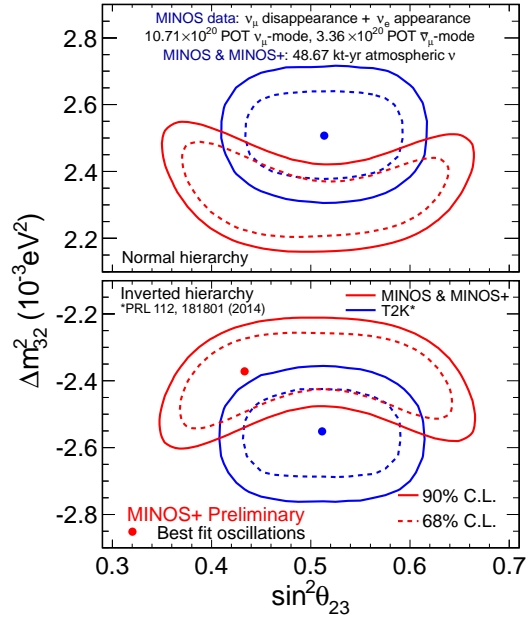}
    \caption{Comparison of contours from the MINOS(+) combined analysis in $\Delta m^{2}_{32}$/$\sin^{2}\theta_{23}$ space with T2K experiment results \cite{t2kres}.   }
\label{fig:t2kcomp}
\end{centering}
\end{figure}

A summary of the numbers of events selected in the combined three-flavour analysis is presented in Table \ref{tab:selevents}. The fit obtains a best fit value for the atmospheric mass squared difference of $\Delta m_{32}^{2}=2.34^{+0.09}_{-0.09} \times 10^{-3}$~eV$^{2}$, and for the mixing angle parameter $\sin^{2}\theta_{23}=0.43^{+0.16}_{-0.04}$ in the case of the normal mass hierarchy. The fit calculates values of $\Delta m_{32}^{2}=2.37^{+0.11}_{-0.07} \times 10^{-3}$~eV$^{2}$ and $\sin^{2}\theta_{23}=0.43^{+0.19}_{-0.05}$ in the case of the inverted mass hierarchy. The 90\% confidence limits (C.L.) for the $\sin^{2}\theta_{23}$ parameter are $0.37<\sin^{2}\theta_{23}<0.64$ in the case of the normal mass hierarchy, and $0.36<\sin^{2}\theta_{23}<0.65$ in the case of the inverted mass hierarchy. The best-fit contours are shown in Fig. \ref{fig:contours} and show a slight preference for the inverted hierarchy and the lower octant of $\theta_{23}$. A comparison of the MINOS(+) results with the T2K experiment results \cite{t2kres} are shown in Fig. \ref{fig:t2kcomp}. The MINOS(+) results represent the most precise measurement of $\Delta m_{32}^{2}$ to date. In the future, it should be possible to combine MINOS(+) and NO$\nu$A data together (when the latter becomes available) to achieve even more precise results and cut into the $\theta_{23}$ octant and neutrino mass hierarchy phase space. 

\begin{table}
\centering
\begin{tabular} {| c | c | c | c |}
\hline
Data Set & No Osc. Pred. & With Osc. Pred. & Observed Events \\
\hline
$\nu_{\mu}$ from $\nu_{\mu}$ beam             & 3201     & 2496    & 2579  \\
$\bar{\nu_{\mu}}$ from $\nu_{\mu}$ beam       & 363      & 319     & 312   \\
Non-fiducial $\nu_{\mu}$ from $\nu_{\mu}$ beam             & 3197   & 2807  & 2911  \\
Atm. contained-vertex $\nu_{\mu}$ and $\bar{\nu_{\mu}}$    & 1414   & 1024  & 1134  \\
Atm. non-fiducial $\mu^{+}$ and $\mu^{-}$      & 732  & 575  & 590  \\
Atm. showers        & 932   &  877  & 899   \\
\hline
\end{tabular}
\caption { Summary of Numbers of Events for the Combined Three-Flavour Analysis.}
\label{tab:selevents}
\end{table} 

\subsection{New MINOS+ Data}
\begin{figure}
\begin{centering}
    \includegraphics[width=0.4\textwidth]{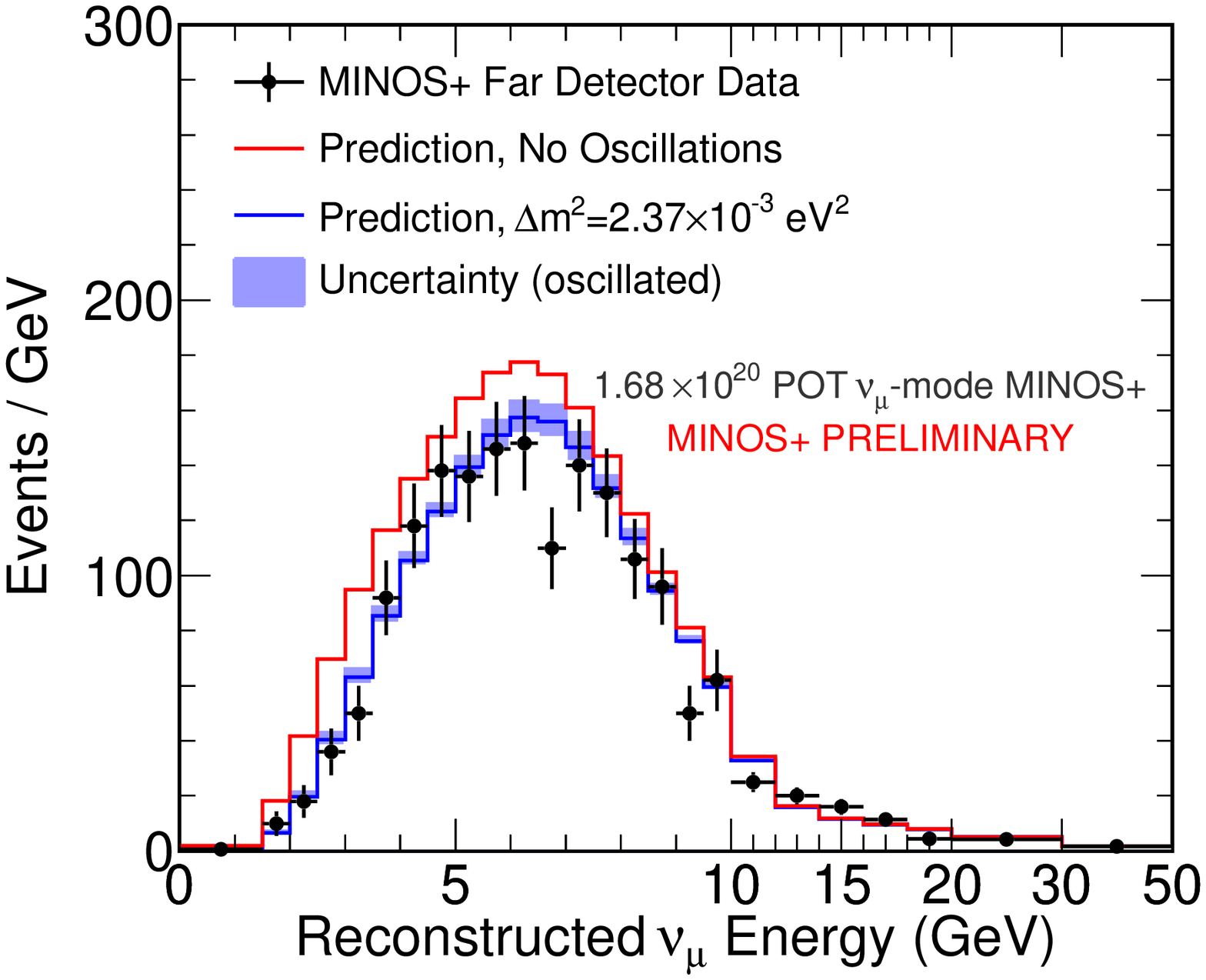}
    \includegraphics[width=0.4\textwidth]{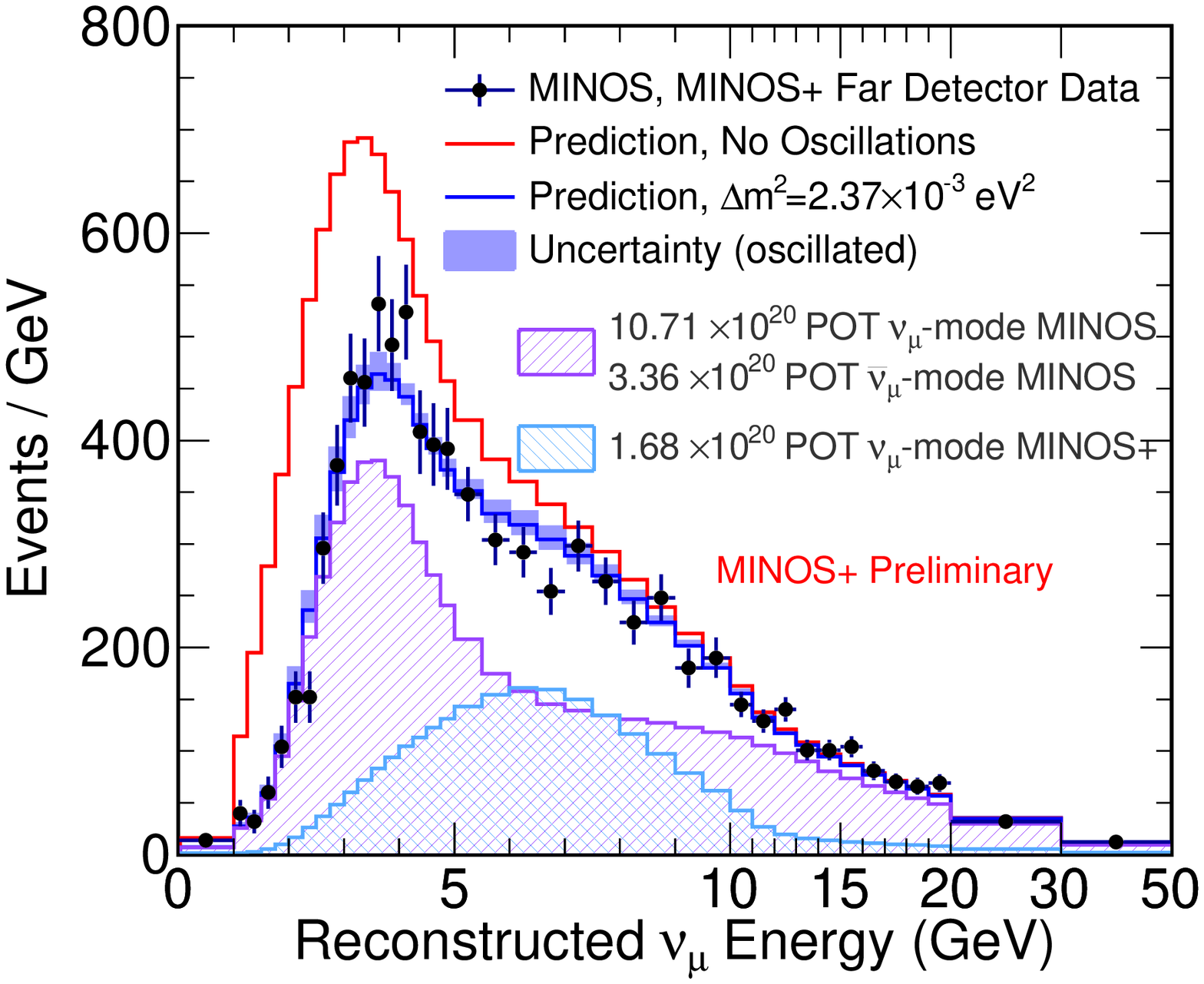}
    \includegraphics[width=0.4\textwidth]{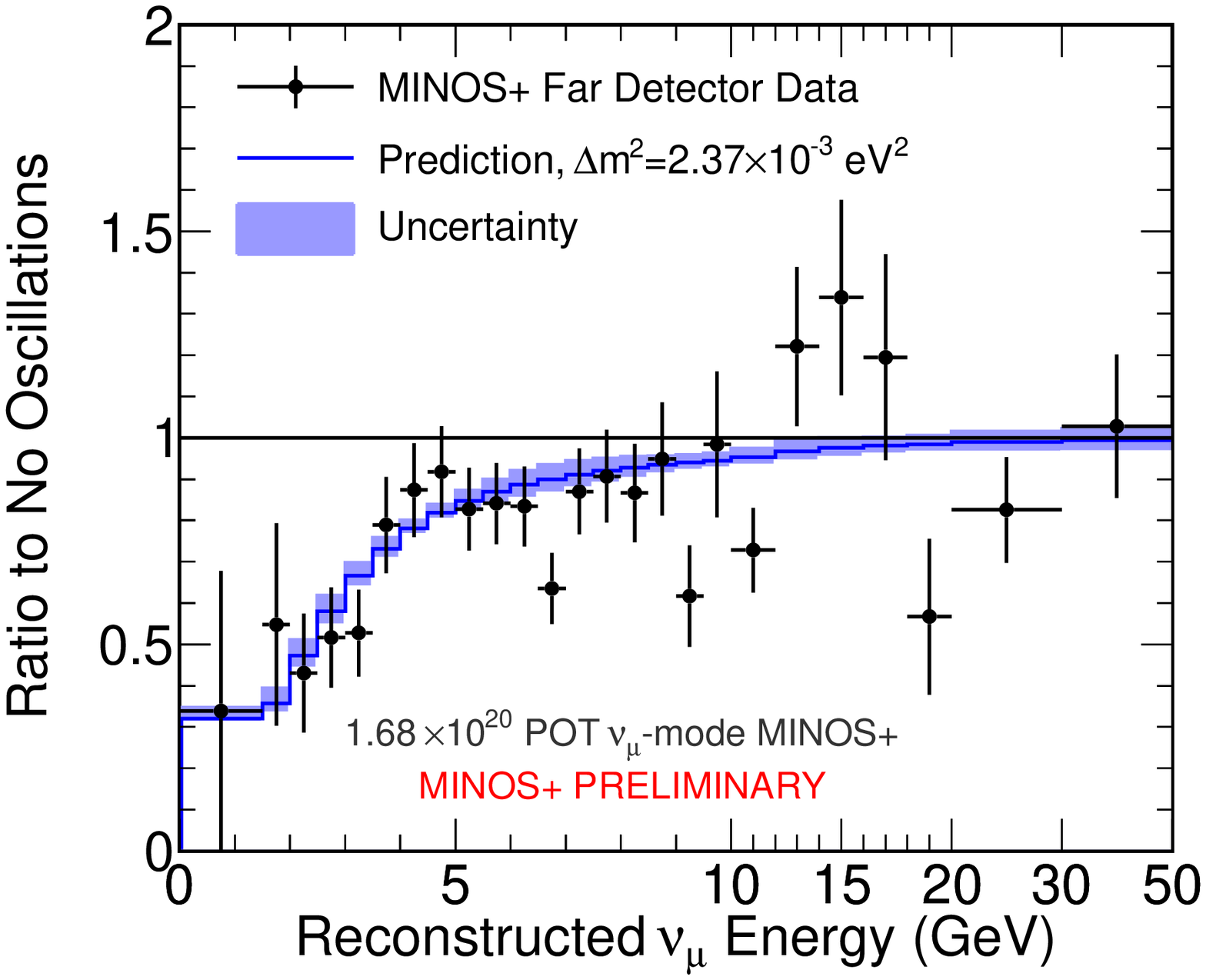}
    \includegraphics[width=0.4\textwidth]{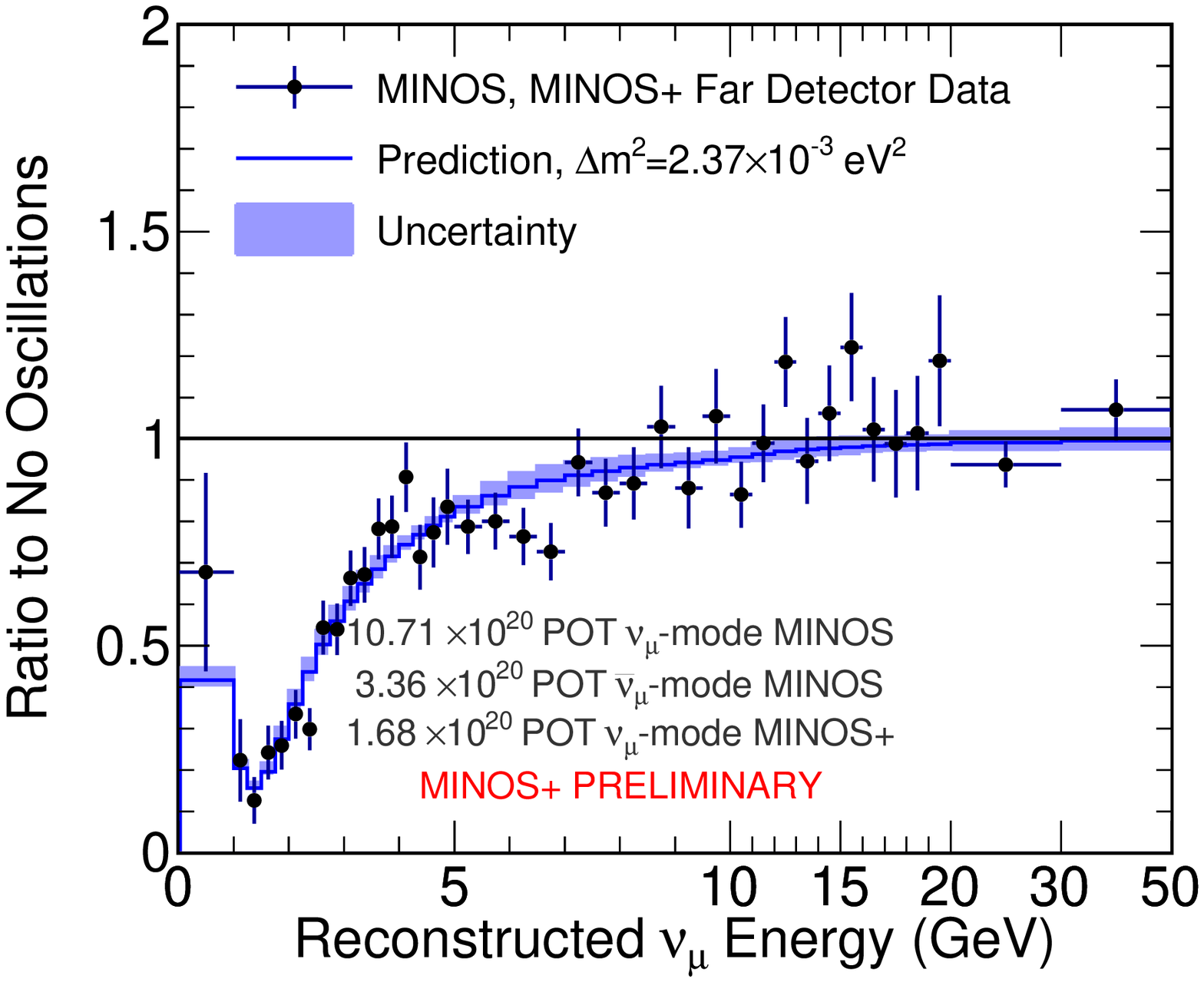}
    \caption{Reconstructed Energy Spectra and Ratios to No Oscillations for MINOS+ Far Detector Selected Events. Predictions are shown for no oscillations and for oscillations at the MINOS(+) best fit. The total one-sigma systematic error is shown for the oscillated prediction. The left panels show the MINOS+ data on its own, and the right panels show the MINOS+ data combined with the MINOS beam data.   }
\label{fig:minosplus}
\end{centering}
\end{figure}

The upgraded NuMI beam restarted on the fourth of September 2013 in ME mode and data from a total beam exposure of 1.68$\times10^{20}$~POT is used for the MINOS+ spectra shown in this paper. The new MINOS+ data are presented in Fig. \ref{fig:minosplus} together with the ratios to no oscillations. The oscillations prediction comes from the best-fit results presented in the previous section. The MINOS+ data are consistent with the MINOS oscillation results. The neutrino data sample contains 1037 events with an oscillated (unoscillated) prediction of 1088 (1255) events. The antineutrino sample contains 48 events, with an oscillated (unoscillated) prediction of 47 (52) events.

\section{MINOS Sterile Neutrino Results}
MINOS uses the simple 3+1 model for its sterile neutrino results. Essentially it looks for distortions from a three-flavour formalism. There are three different regimes in which MINOS is sensitive to a fourth sterile neutrino. For small $\Delta m_{43}^{2}\lesssim$0.1~eV$^{2}$, in the ``slow'' sterile oscillations regime, spectrum disortions are expected well above the standard oscillations maximum in the FD, with no effect in the ND. For medium values of 0.1~eV$^{2}\lesssim \Delta m_{43}^{2}\lesssim 1$~eV$^{2}$, in the ``intermediate'' regime, sterile oscillations at the FD become fast and average out and result in a counting experiment, still without effect in the ND. For large $\Delta m_{43}^{2}\gtrsim$1~eV$^{2}$ however, in the ``rapid'' sterile oscillations regime, there are rapid oscillations at the FD which average out, and in addition there is an oscillations effect in the ND which affects the extrapolation to the FD.

In order to account for ND distortions of the spectrum due to sterile neutrino oscillations, MINOS looks at the Far/Near ratio for both the CC and NC events and fits the oscillated F/N ratio directly to the F/N data ratio using a $\chi^{2}$ function. It fits for several parameters: $\Delta m_{32}^{2}$, $\theta_{23}$, $\Delta m_{43}^{2}$, $\theta_{24}$ and $\theta_{34}$. Systematic uncertainties are reassessed, including beam systematic uncertainties which become more important due to the possible ND oscillations for large values of $\Delta m_{43}^{2}$. Covariance matrices are used to include a total of 26 different systematic uncertainties in the fit, and the Feldman-Cousins approach is used to correct the log-likelihood surfaces \cite{FC}. For the results shown here, the full MINOS LE neutrino-dominated beam mode data set of 10.56 $\times 10^{20}$~POT is used.

\begin{figure}
\begin{centering}
    \includegraphics[width=0.4\textwidth]{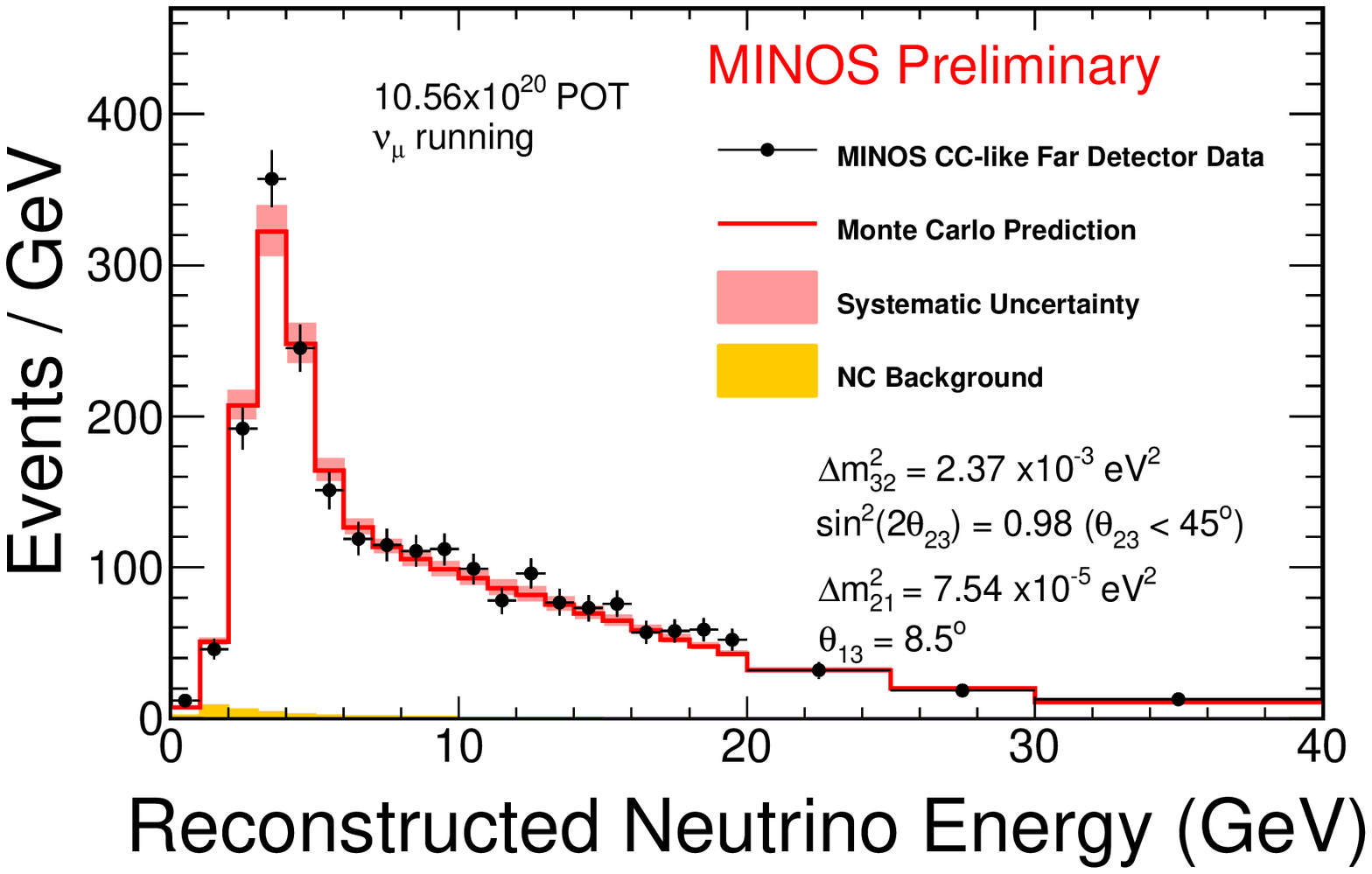}
    \includegraphics[width=0.4\textwidth]{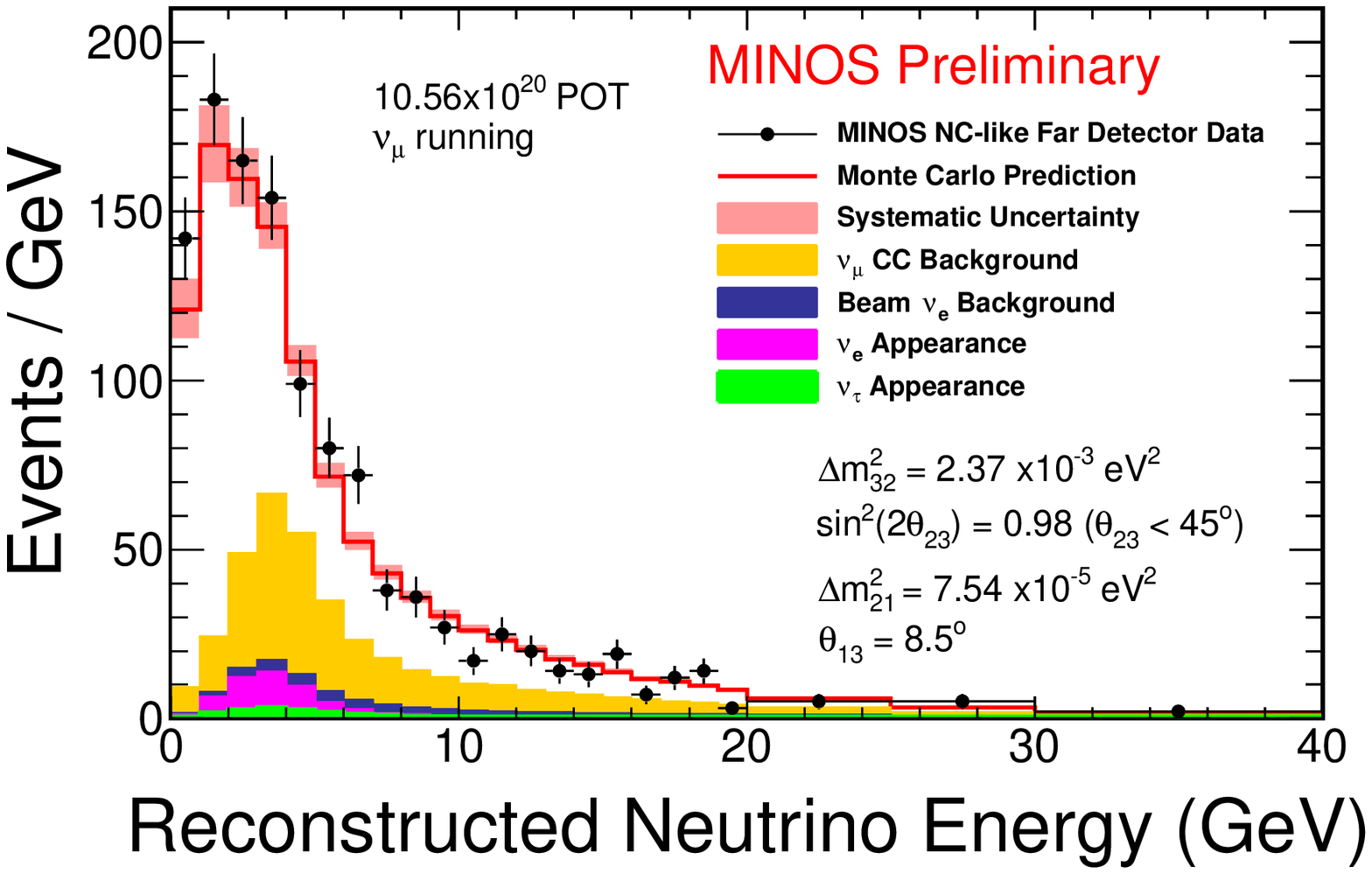}
    \includegraphics[width=0.4\textwidth]{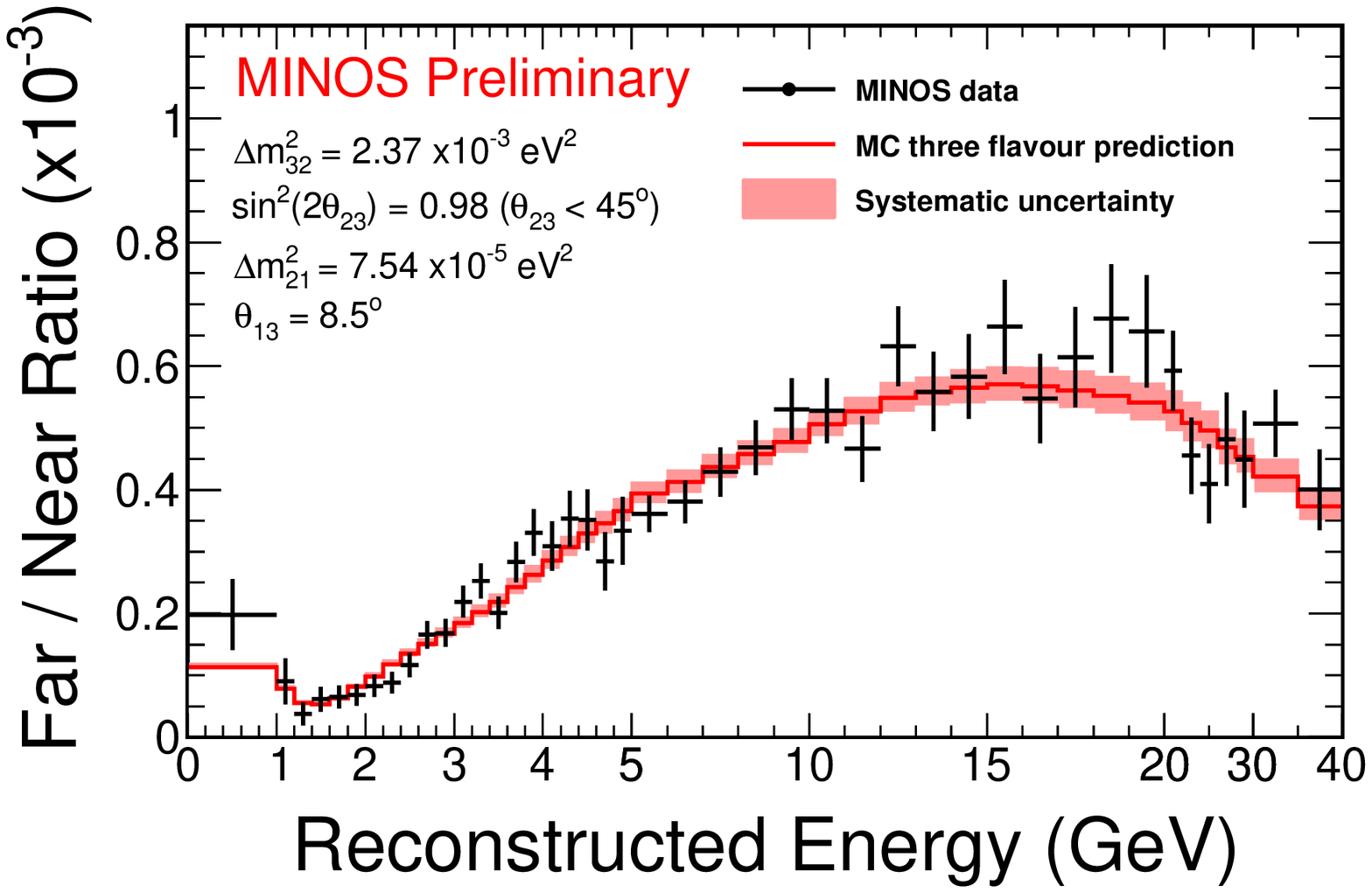}
    \includegraphics[width=0.4\textwidth]{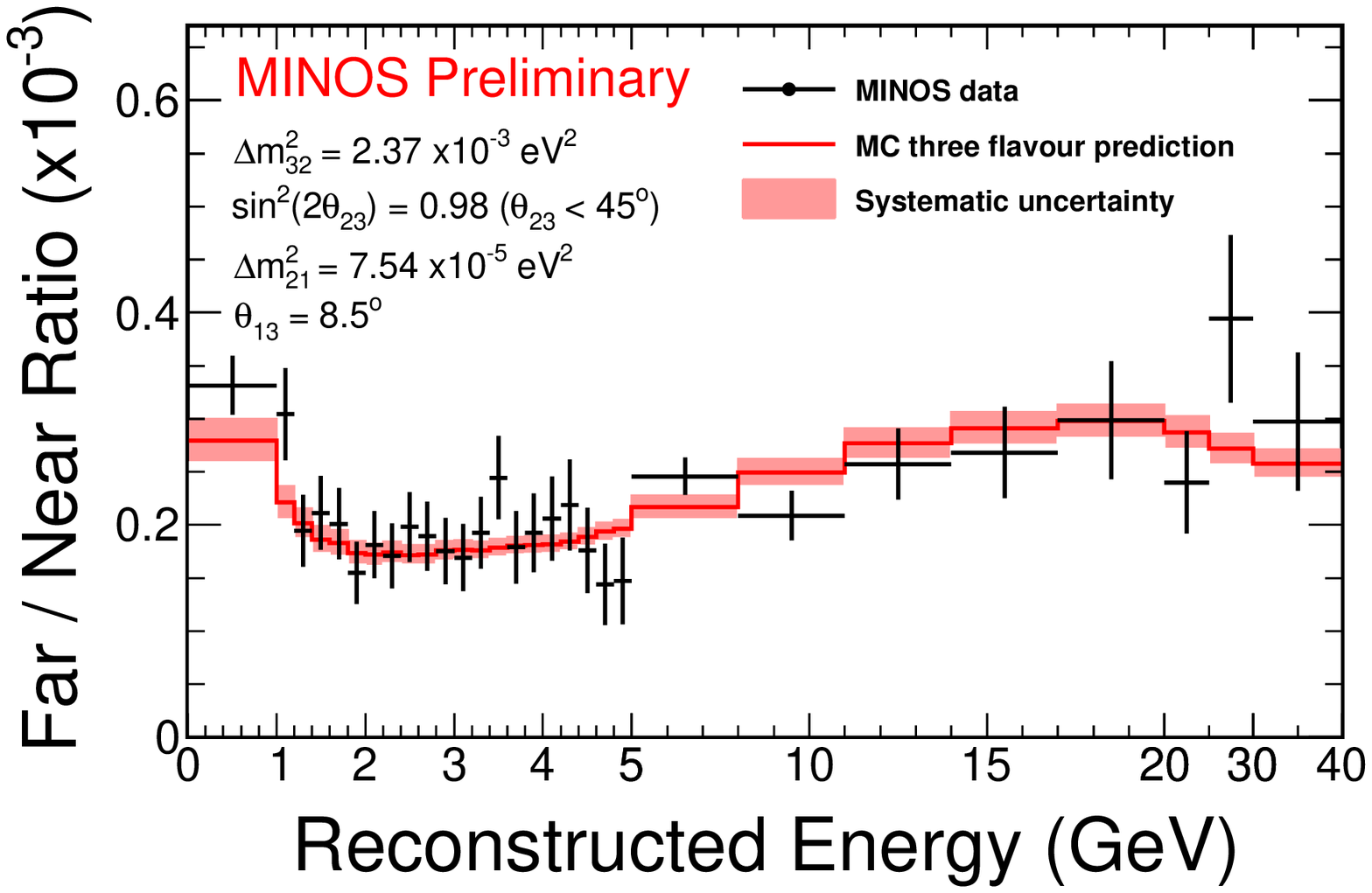}
    \caption{Far Detector Spectra Comparison for the MINOS Sterile Analysis. This graph shows a comparison of the FD Charge Current and Neutral Current event data spectra in black to the MINOS three-flavour best-fit prediction in red. The red bands represent systematic uncertainties on the prediction. Various backgrounds are shown in colour. The Far/Near ratios for both data and prediction are also shown for both samples. The prediction is fit to the data Far/Near ratios to determine the oscillation parameters.  }
\label{fig:sterilespectra}
\end{centering}
\end{figure}

\begin{figure}
\begin{centering}
    \includegraphics[width=0.6\textwidth]{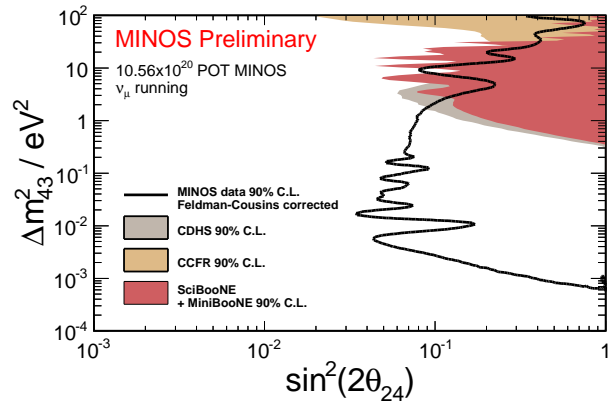}
    \caption{Disappearance Exclusion for Sterile Neutrino Oscillations. The 90\% C.L. exclusion limit in $\Delta m_{43}^{2}$ and $\sin^{2}(2\theta_{24})$ space is shown in black for the total MINOS neutrino exposure of 10.56$\times 10^{20}$~POT. Other experiment results are shown in colour. MINOS sets the world's best limit over several orders of magnitude in $\Delta m_{43}^{2}$.   }
\label{fig:sterileres}
\end{centering}
\end{figure}

Fig. \ref{fig:sterilespectra} shows a comparison of the FD data with the standard three-flavour prediction for the full MINOS LE neutrino-mode beam sample, including the corresponding Far/Near ratios. 2712 $\nu_{\mu}$-CC-like events, and 1221 NC-like events, are observed in the FD. After the final fit of the Far/Near predictions to the Far/Near data, MINOS gives the world's strongest constraint on muon neutrino to sterile neutrino oscillations for $\Delta m_{43}^{2}\lesssim$1~eV$^{2}$. The corresponding contours can be seen in Fig. \ref{fig:sterileres}. Moving forward, additional neutrino data from MINOS+, and also data in antineutrino-enhanced beam mode will provide even better limits for the sterile neutrino disappearance analysis. 

\section{Summary}
The MINOS experiment has been reborn as the new MINOS+ experiment in the upgraded ME NuMI beam. It started taking data in September 2013 and has already contributed to the standard oscillations three-flavour analysis with additional atmospheric neutrinos data. This combined three-flavour analysis obtains the atmospheric parameter best-fit point of $\Delta m_{32}^{2}=2.37^{+0.11}_{-0.07} \times 10^{-3}$~eV$^{2}$ and $\sin^{2}\theta_{23}=0.43^{+0.19}_{-0.05}$ in the inverted hierarchy phase space. The data shows a slight preference for the inverted hierarchy and the lower octant of $\theta_{23}$. A first look at the new MINOS+ beam data is shown and is consistent with the updated results from the standard oscillations analysis. Finally, new MINOS results for the search for sterile neutrinos are presented and new constraints are placed on the possible phase-space accessible to potential sterile neutrinos using neutrino disappearance.  

\section{Acknowledgements}
This work was supported by the US DOE, the United Kingdom STFC, the US NSF, the state and University of Minnesota, and Brazil's FAPESP, CNPq and CAPES. Fermilab is operated by Fermi Research Alliance, LLC under Contract No. De-AC02-07CH11359 with the United States Department of Energy.

\end{document}